%% file: Waterfall.tex
\documentclass[10pt,preprint,letter]{sigplanconf}
\RequirePackage[utf8]{inputenc}

\usepackage{ifthen}
\newboolean{showcomments}

\input{header.tex}

\newcommand{\W}{Waterfall\xspace}
\newcommand{\B}{Benzo\xspace}

\newcommand\cb[1]{\nbc{CB}{#1}{magenta}}

\begin{document}

\title{\W: Primitives Generation on the Fly}

\authorinfo{Guido Chari \and Diego Garbervetsky}
    {Departamento de Computación, FCEyN, UBA}
    {\url{http://lafhis.dc.uba.ar}}
\authorinfo{Camilo Bruni \and Marcus Denker \and Stéphane Ducasse}
    {RMoD, INRIA Lille - Nord Europe, France}
    {\url{http://rmod.lille.inria.fr}}

\maketitle



\begin{abstract}
Modern languages are typically supported by managed runtimes (Virtual Machines). 
Since VMs have to deal with many concepts such as memory management, abstract execution model and scheduling, they tend to be very complex.
Additionally, VMs have to meet strong performance requirements. 

This demand of performance is one of the main reasons why many VMs are built statically. 
Thus, design decisions are frozen at compile time preventing changes at runtime.
One clear example is the impossibility to dynamically adapt or change primitives of the VM once it has been compiled.
  
In this work we present a toolchain that allows for altering and configuring components such as primitives and plug-ins at runtime. 
The main contribution is \W, a dynamic and reflective translator from Slang, a restricted subset of \ST, to native code.
\W generates primitives on demand and executes them on the fly. 
We validate our approach by implementing dynamic primitive modification and runtime customization of VM plug-ins.
\end{abstract}

\section{Introduction}
Modern high-level languages usually rely on runtime systems (Virtual Machines) which provide abstract execution models that promote portability, automatic memory management and enforce certain security properties. 
VMs are complex pieces of software that historically, for efficiency reasons, were developed in static low-level languages. 
Over the last years a new branch of VM implementations written in high-level languages appeared~\cite{Folla}.
By relying on better abstractions the VMs would be easier to develop, debug and modify. 
The downside of the gain in abstraction is a further separation from the low-level execution model of the underlying hardware. 
It requires substantial efforts to make VMs written in high-level languages efficient~\cite{Rigo06a}.

Self-hosted VMs~\cite{Chev11a} are written in the same language they support.
A common way of tackling them was to restrict the actual semantics for the code targeted to the VM construction. 
This is the path taken in many implementations like Squeak~\cite{Inga97a} for~\ST and PyPy~\cite{Rigo06a} for Python.
The code of these VMs is written in a subset of the language they provide.
This subset typically has the same syntax, but the semantics are restricted in order to statically bind everything at  compile time. 
Essentially, this prevents polymorphism and dynamic method dispatch which are substantial features of high-level host languages.
A quite similar approach is observed in popular research VMs like Jikes~\cite{Alpe05a} or Jalapeño ~\cite{Alpe99a} for Java.

VMs written in high-level languages have improved considerably and it has become a prolific area~\cite{Wimm13a,Chev11a}.
One promising technique is by promoting low-level high-level programming frameworks~\cite{Fram09a} which aims at solving the problem of the abstraction mismatch of metacircular or self-hosted VMs. 
With these elaborate tools and others like runtime code specialization the performance penalty can be considerably mitigated.


However we identified that most of these VMs still suffer from an important limitation:
\\

\emph{It is not possible to change or configure many of the most important design decision at runtime. Key components are frozen at compile time and hidden at runtime}\\

One example of this limitation are the primitive methods. They are used by VMs to perform basic operations such as arithmetic operations and object allocation or to accomplish performance demanding task more efficiently~\cite{Gold83a}. 
They live at VM-level and are statically bound at compile time. 
In certain cases a developer may be interested in creating or adapting a primitive, for instance for instrumenting an application. 
In some languages this can be achieved by exploiting reflective capabilities \cite{Roet07b}.
But the overhead imposed by actual implementations of reflection make them impractical for performance demanding tasks~\cite{Male96a}.

Similarly, some VMs support the concept of plug-ins to enable the efficient implementation of recurrent general operations (e.g, file access, floating point operations, compression, etc).
Plug-ins also provide some features that can not be implemented fully at language-side.
Moreover, only after releasing, developers may identify a hotspot in their application and improve its performance by providing a plug-in for it.
Unfortunately, in many cases this implies that every user of the application will need to recompile its VM in order to update the plug-in.
Other option is to ship the plug-in as a library and depend on the user's platform compilers or dynamic linkers.
Clearly, giving end-users responsibilities, or worse, requiring them to update their VM every time a new plug-in is released is undesirable. 

To overcome this limitations we conceived a toolchain which solves the aforementioned problems. 
Its main component is \W, a runtime translator for code in a syntax equal the the language the VM supports. 
We take the approach of high-level low-level programming one step further and use it dynamically at runtime to change VM behavior.
Concretely, \W enables to dynamically modify primitives and plug-ins from language-side, outperforming  existing approaches that are purely reflective. 
We use a single-language approach instead of using Foreign Function Interfaces (FFI)~\cite{Lian99a,Mill03a} to access external libraries.
This enables to debug, inspect and change the code at runtime with the same tools the developers use for their general purpose tasks. 
Exploiting \W capabilities, developers and experienced users are able to tune their VMs at runtime without the need of external tools.

The contributions of this paper are:
\begin{itemize}
	\item A toolchain, written in the language of the VM, that enables to compile and activate code at runtime.
	\item A proof of concept that uses the toolchain for adapting VM behavior at runtime, without the need of a system restart, considerably outperforming the reflective solutions. 
	\item An empirical validation demonstrating the approach is feasible and the penalty in terms of performance is reasonable penalty.     
\end{itemize}

\section{Statically Defined Primitives}
\seclabel{problems}
One of the main components of VMs are primitives. 
Primitive routines \ugh{provide} the runtime with the capability to perform essential operations that the language could not supply by its own. 
For instance, create objects or provide access to low-level structures. 
On the other hand, they are also used to optimize some critical bottlenecks~\cite[ch.3, p.52]{Gold83a}

Each time the runtime activates a method that is a primitive, it swaps the execution mode. 
The VM, instead of interpreting bytecodes or performing message dispatching, directly executes the binary code of the primitive.
Primitives typically are already statically compiled, further optimized and in general they are written in the same language used to implement the VM.
Changing or extending primitives is profitable in the case where core features of a language needs to be inspected or modified.
For instance, if a VM  accesses its instance variables through a primitive, immutability could be easily and efficiently achieved by changing that primitive.

The problem is that primitives are deeply coupled to the VM building process which is a complex and time-consuming task. 
Moreover, changing primitives may not be possible since it requires access to the VM source code, which is not always open.
Even in that case, primitives are written on a different abstraction level than the high-level language the VM supports. 
This is a complex barrier for the common developer of the host language since he needs to work in two abstraction levels at the same time and deal with different development environments and tools~\cite{Fram09a}. 
As a consequence it is observed that the set of available primitives is statically defined and frozen at compile time, making them difficult or even impossible to change. 

Self-hosted VMs try to overcome this two-language problem by implementing the VM using a language with a similar syntax as the host language, mitigating several of the problems mentioned previously. 
But, even though the syntax is usually similar, or even equal, the developer still has to be aware that the semantics are different from the host language. 
For example, in Squeak and PyPy the VM is implemented in a language that is almost identical to the host language but eventually the semantics are reduced to C-like expressions. 

Another example similar to primitives in high-level dynamic environments is the \PH VM~\cite{Pharo,Blac09a} plug-in infrastructure. 
\PH provides a large number of plug-ins tailored towards specific but heterogeneous tasks such as algebraic matrix operations, floating point precision computations, file management, etc. 
All these tasks have in common that they are intensively used for repetitive and performance demanding tasks that are not fulfilled by the standard runtime execution model. 
In essence, plug-ins consist of a set of primitives isolated on separated modules.   

Plug-ins are, in many cases, built and deployed with the VM.
This is unpleasant and involves exactly the same limitations as static primitives in a larger scale.
The other scenario is to build them as dynamic linked libraries (DLLs), like modules outside VM, and link them dynamically.
In an case, plug-ins imply bigger binary footprints, more complexity and overhead for deployment.
Plug-ins force users to work with VM binaries with a large amount of statically built code that they perhaps never use.
Concerning runtime adaption, the only option to change and modify plug-ins is by using the second approach (DLLs) and by loading them dynamically with external OS tools, possibly after unloading a previous version.
Relying on external compilers brings its own set of problems.
Additionally to the given CPU architecture difference, compilers and linkers differ on each operating system.
\\\\
Below we summarize the limitations of current VMs, concerning runtime adaptation, that we want to address: 
\begin{enumerate}
	\item Primitives are statically defined and frozen at compiled time. 
	\item Primitives can not be efficiently changed at runtime.
	\item It is complex, heavy-weight and error prone to extend or upgrade VM modules at runtime.
\end{enumerate}

\section{Context}
\ST is a good representative of a dynamic, object-oriented and reflective high-level language.
We choose \PH\footnote{\url{http://pharo.org}}~\cite{Pharo,Blac09a}, an open-source \ST-inspired environment, for developing the proof of concept used to validate our approach.
\PH features a self-hosted VM with primitives accessible at language-side. 
Moreover, the VM  provides a powerful plug-in infrastructure with a simple interface, enabling its extension  with efficient low-level functionalities, which works very similar to primitives. 

In the rest of this section we provide context for understanding the \W approach presented on next section.

\subsection{\PH VM}
The VM for \PH is the Cog VM~\cite{Mira11a}, which has a unique construction process that was inherited from the Squeak VM~\cite{Inga97a}. 
The language under which Cog is developed is a subset of \ST known as Slang. 
This has the advantage that Slang programs can be managed and explored in the same way as any other code from the host language. 
The main difference is that Slang code is not executed by the runtime.
Instead there is a compiler that translates Slang to C, wherefrom a standard C compiler takes over.
This is why Slang basically has the same syntax as \ST but it is semantically constrained to expressions that can be resolved statically at code generation time.
Hence Slang’s semantics are closer to C than to \ST.

In Cog, primitives are defined at VM-side using Slang and they are frozen at compile time.
Following a similar approach as primitives the Cog VM can be customized with plug-ins which are also written in Slang and use the same compilation strategy already explained. 
Plug-ins can be seen as modules that encapsulate a set of particular primitives.
One of the main differences is that plug-ins can be compiled in external files, like independent libraries, and then linked to the VM at runtime.

\subsection{\B Framework}
\W relies on a framework called \B~\cite{Benzo13} that provides dynamic high-level low-level programming techniques.
\B provides a language-side assembler, a dynamic code generator and a set of generic primitives for activating native code from language-side.

Instead of building a separate infrastructure for generating and activating native code we rely on this generic framework.
\B is based on 5 generic primitives to activate native code and access VM symbols.
\B uses its own language-side assembler to generate native code without external help.
Once the native code is ready \B uses a dedicated primitive to active the code.
This very basic interface is enough to open doors for new language-side implementations of typical VM-level tools.
\B is already successfully used in production for a language-side FFI implementation~\cite{Brun13a}.




\section{\W Compiler}
\label{waterfall}
\W's main task is to translate Slang to native code at runtime.
The compiler was designed as a chain of transformations that receives as input a high-level representation of code (Slang) and returns as output a lower-level one (binary code) going through different intermediate representations in the process. 

\subsection{General Architecture} 
\begin{figure}
\begin{center}
\includegraphics[width=\linewidth]{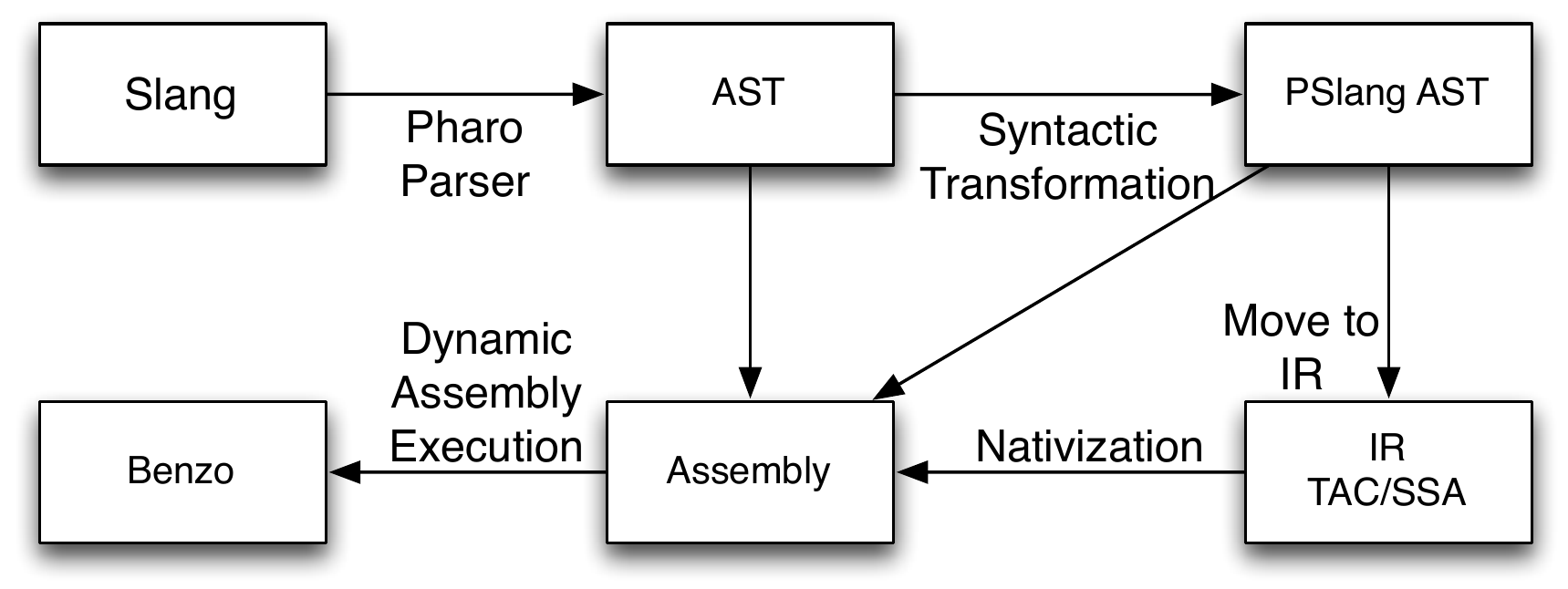}
\caption{All \W stages for Slang to native compilation.}
\label{fig:ejemplo}
\end{center}
\end{figure}

A high-level graphical description of the processes involved in generating native code is shown in Figure \ref{fig:ejemplo}.
The \W toolchain begins by parsing Slang to an Abstract Syntax Tree (AST) representation. 
Then the AST representation is translated to native code enforcing C-like Slang semantics.
\W generates assembler instructions at language-side and executes them using the \B framework.
\W can also translate the AST into an intermediate representation (IR) featuring an hybrid IR between Three Address Code (TAC) and Static Single Assignment form (SSA) as a pre-step before native code generation. 

Even though Slang is syntactically similar to \ST it is actually closer to C. 
It is completely bound at compile time and most of their constructs can be directly translated to C. Considering that fact, the binary code that \W generates follows similar code conventions used by  C compilers.   
Concretely, Slang variables and arguments are pushed on the native stack and Slang messages are treated very similar to C function calls. 
\W also sees every variable just as a word in memory leaving its semantics interpretation (type) to the developers. 

\begin{figure}[H]
\begin{center}
	\includegraphics[width=\linewidth, height=120px]{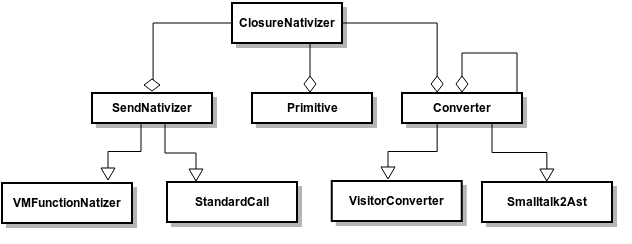} 
	\caption{\W high-level class diagram}
	\label{fig:classdiag}
\end{center}
\end{figure}

Figure~\ref{fig:classdiag} exposes a high-level diagram of the core classes involved in the translation. 
\W's main component is the \textbf{ClosureNativizer} that has as collaborators: a set of \textbf{SendNativizers}, a static set of \textbf{Primitives} of the language and a chain of \textbf{Converters}. 

\paragraph{Primitives.}
Certain operations at Slang-level can not be expressed as calls to other Slang methods and thus are defined as assembler templates. 
Much like in the normal \PH environment, Slang relies on certain primitive operations that can not be expressed at the same level.
For instance, algebraic operations such as $+$, $-$, bit manipulation or direct memory access operations have to be treated differently.
Instead of generating their native code from a Slang method we inline an already nativized template (instance of the \textbf{Primitive} class).

\paragraph{SendNativizer Hierarchy.}
The \textbf{SendNativizer} objects have the responsibility of handling the different type of calls (calling convention) which we describe in more detail in Section~\secref{slang-ast-to-native}.
 
\paragraph{Converter Hierarchy.}
Converter objects essentially receive as input a representation of the code, apply transformations and emit a modified output.
A chain of these transformations defines the compilation phases that \W will actually execute. 
\textbf{VisitorConverter} is the class that represents the behavior for walking through nodes in a visitor fashion. 
All transformations that traverse ASTs doing something different for each kind of tree node are implemented as visitors and collaborate with a \textbf{VisitorConverter}.
With this architecture it is quite simple to define new conversions such as optimizations or further IRs, and link them to the chain.
As an  example of the reuse of standard environment tools, \textbf{Smalltalk2AST} converter is instantiated with the \PH standard parser. The converter is named Smalltalk since their syntax is the same, but \W uses it for translating Slang.

\subsection{Replacing Primitives with Generated Code}
Once \W generated the code, there still remains the step of actually installing and executing it at runtime.
In the case of changing a primitive, the generated code must be executed instead of the original static primitive.
Below we describe a summary of the main steps involved in this task: 

\begin{enumerate}
	\item Exploiting reflective properties of the environment with meta programming techniques, \W identifies whenever a plug-in or primitive is called. 
	In the \PH environment these methods are identified by a special \textbf{pragma} as first statement.

	\item The pragma is removed and instead \W installs code that invokes the generation of a new primitive on demand from Slang code.

	\item Finally, \W relies on \B for executing the native code. Subsequent calls to the same primitive function will not execute step 2, avoiding the compilation overhead.
\end{enumerate}

\subsection{Replacing Plug-ins with Generated Code} 
There is only a marginal difference when using \W for generating plug-ins since plug-ins are in some way modules encapsulating a set of related primitives. They are more general.
For dealing with them there exist a class \textbf{PluginNativizer} whose instances receive as input (collaborators) a set of methods (primitives) from a plug-in.
Additionally \textbf{PluginNativizer} knows the class where the new dynamically nativized primitives should be installed.
This nativizer is responsible for calling the translator on each of the methods and installing the code generated as new primitives on the corresponding class. 

With this approach it is quite simple to mark a plug-in from language-side as dirty for later recompilation.
Concretely, marking it as dirty implies to trigger a call to the nativizer with only the changed method as input.
Since the interface described above is written at language-side, and \ST is a reflective language, all this behavior can be dynamically activated.
Plug-ins developers can also define their plug-ins as language functions that trigger the nativizer whenever called for the first time.
It is also easy to make changes to a plug-in or create a different version and let the user decide to use the one that better fits his requirements.

\subsection{Compilation Steps}
In the rest of this section we describe the standard steps of transformations (\textbf{Converters}) needed for translating Slang code to native instructions.

\subsubsection{Transforming Slang to AST} 
This step benefits from the cooperation of high-level tools at language-side provided by the \PH environment.
Concretely, since Slang is a subset of \ST, \W uses the standard \ST parser to generate the AST.
The parser itself is written in \ST and resides at language-side, which makes it an easy target for debugging and possible extensions.
Hence \W does not need a special parser.

\subsubsection{Computing the Set of Reachable Methods}
Before executing a Slang method \W translates its code into machine instructions. 
However looking at an isolated method itself is not enough. 
Slang programs can be decomposed in several methods so  
\W also has to translate all the methods reachable for all possible executions starting at the method under translation. 
Essentially, every potentially executable method has to be nativized.
However it must remain clear that \W currently does not support polymorphism, only modularization.

\subsubsection{Transforming AST to Native Code}
\seclabel{slang-ast-to-native}
The actual nativization step consists of a visitor that receives an AST as input and returns a stream of bytes which then is loaded into memory for execution.
For abstracting from the binary code generation, we use an already developed abstract assembler called ASMJit\footnote{ASMJit: \url{https://code.google.com/p/asmjit/}} that is an independent tool inside the \B framework.    

The Slang AST has few kind of nodes. Each node is translated into a set of native instructions.  
As an example, we briefly explain how to translate nodes corresponding to variable references (more details about the translation of other nodes in section~\ref{sec:implementation}).
The visitor checks what kind of variable is referenced (i.e., temporary, argument, global, low-level symbol)  and it finds its memory address: for temporaries and arguments it will be a native stack location, for VM globals the position where they were loaded in memory.
Finally, the translator interacts with ASMJit for pushing the gathered memory address into the stack.  


\subsubsection{Dynamically Executing Generated Native Code} 
After generating the native instructions there still reamins the task of activating the native code and passing it down to the VM.
This is not directly possible since in general VMs strongly separate the language from the low-level environment.
Due to this barrier it is simpler to ensure portability and security properties for a VM.
However, in our case this poses a limitation since we want to dynamically execute instructions at VM-level which were created at language-side.
As already explained, \W is supported by \B framework for overcoming this limitation.

\section{Implementation Details} \label{sec:implementation}
The main goal of \W  is to provide support for changing low-level (VM) behavior at runtime.
Due to the existing restrictions of the VM we apply certain simplifications or rely on external tools to accomplish our goal.

\subsection{Resolving External Symbols} 
Code written by developers may reference other code or data that already exists in the managed runtime, like global variables or VM functions.
In these cases, the nativization procedure has to find the actual addresses in memory where these static references lie and inline them into the generated native code.
All this is necessary because Slang eventually resides at VM-level when executed and thus it is not possible to access global objects directly.
While executing a primitive it is not possible to interact with the high-level environment. 
For instance accessing an object using the standard \ST way by sending a message would cause recursion problems.

To access the position of VM internal symbols referenced in Slang methods, \W relies on an existing \B API to interface with C libraries which is based on~\textbf{dlsym}. 
We developed also a parser of the \textbf{nm} Unix command for gathering the positions not accessible to \textbf{dlsym} in the VM binary. 
As a consequence of the external tools selected, the complete functionality of the compiler can be only obtained on Unix platforms. 
However, it should not be difficult to develop a parser for a  tool similar to \textbf{nm} on Windows or other platforms.

\subsection{Special Parameters}
\PH primitives receive their arguments on the stack. 
The VM is responsible for pushing them right before calling the primitive.
Low-level primitive code special has to be careful when accessing these arguments.
The Cog VM uses a moving Garbage Collector (GC) which requires careful access to high-level objects when in VM-level code.
The GC is not aware of the C stack. Therefore, if a GC pass happens during primitive activation, \W pointers to language-side objects held on the C stack are not updated resulting in dangling pointers that would cause severe troubles.
\W simplifies the access to high-level objects by using a single statically known position for all parameters.
This way only a single memory address has to be registered at the garbage collector for not moving it.
Since the VM is single threaded and no two primitives can be executed at the same time it is safe to rely on a single global argument position.

\subsection{Slang Purification}
The complete Slang language supports special syntax for inlining C expressions.
That means actually that it is at least as expressible as C.
It also allows for other expression such as type \textbf{pragmas} which contain type information and are directly translated to C types. 
\W currently does not fully support these two support special Slang expressions.
However to validate our approach we reimplement these two features in a simple fashion.

In the case of types there exists a special converter that walks over the \textbf{pragmas} of the nodes, parses them and finally assigns basic type information to the variables that were reified during the previous transformations. 
The converter also marks if their arguments are used as value or as reference.
Concerning the inlined C macros, we realized that they were mostly used to call external functions so we decided to provide a special language construction only for that cases.

We finally provide a special converter that performs simple string substitutions for the Slang features we implement differently.

\subsection{Managing Stack Frames}
The most complex node of the AST (Section~\secref{slang-ast-to-native}) to deal with by \W is the one for messages in \ST which is reduced to function calls in Slang.
The generation of native code in this case implies defining a calling convention for argument passing and register preserving (similar to C in this case): pushing correctly the arguments, preserving the needed registers, calling to the right place and resuming control at return. 
For each function call, a \textit{Context} object is instantiated which represents the stack frame and has a pointer to its parent. 
This context is responsible for determining the actual position on the stack for every reference to a variable. Since \W allows blocks (lambda functions), it must manage very carefully the context stack since a variable reference could be in another context far from the current one.
Finally, as an example of the different calling conventions, calling VM functions implies sticking to a C ABI, whereas for Slang internal calls there is a receiver of the function (implicit parameter) that is always pushed on the stack.


\section{Validation}
We present two case studies: an essential primitive and a language-side plug-in untied from the VM building process. 
The first experiment validates that our solution efficiently addresses the first two limitations  identified in Section~\secref{problems} concerning primitives.  
The second experiment shows how \W overcomes the limitation regarding language-side VM extensions. We run all the benchmarks for this paper with the SMark\footnote{\url{http://smalltalkhub.com/\#!/~StefanMarr/SMark}} benchmarking tool. On each benchmark we measure 50 runs and take the average time. 
 
\subsection{Essential Primitives Instrumentation}
We distinguish two types of primitives: essential and non-essential. 
Essential primitives are required for the bootstrap and vital operations of the language, such as creating a new object or activating a block. 
Such primitives can not be easily implemented at language-side.
The second category of primitives are mainly used for optimization purposes and could be replaced by language-side code.

In this particular case study, we focus on essential primitives.
Instrumentation of essential primitives is an error-prone task falling in many cases in non-termination due to recursive loops.  
A difficult candidate is the \textbf{basicNew} primitive, which is responsible for instantiating new objects.
Even a very simple instrumentation task such as printing the address in memory of the created  object is problematic.
If during the printing process another object is created, the very same instrumented \textbf{basicNew} primitive would be triggered.

Using reflective techniques it is possible to avoid this loop. 
Essentially one would have to inspect the current stack for previous activations of the modified primitive before using it.
If the primitive has been used before, the code jumps to the original unmodified primitive. 
Thus, breaking the recursive loop. However this approach imposes a considerable overhead.

\begin{table*}[t]
    \centering
    \begin{tabular}{llcc}
                                           & Average Time                     & Relative Time      & w.r.t Waterfall Instr.\\\midrule
        Unmodified                         & \ttt{ 0.28} $\pm$ \ttt{0.16} ms  & $1.0 \times$       & $-$\\
        Unsafe reflective instrumentation  & \ttt{21.80} $\pm$ \ttt{0.33} ms  & $\approx 78\times$ & $\approx 2,8 \times$\\
        Secure reflective instrumentation  & \ttt{27.72} $\pm$ \ttt{0.40} ms  & $\approx 99\times$ & $\approx 3.6 \times$\\

        \W-based instrumentation           & \ttt{ 7.72} $\pm$ \ttt{0.27} ms  & $\approx 28\times$ & $1.0 \times$\\
        \W-based unmodified                & \ttt{ 7.08} $\pm$ \ttt{0.23} ms  & $\approx 25\times$ & $-$\\
    \end{tabular}
    \caption{Slowdowns comparison for instrumentations of the  essential primitive \textbf{basicNew}.}
    \label{benchmark-basicnew}
\end{table*}

\subsubsection{Experiment}
We present the code of two approaches for instrumenting the object creation primitive: a pure language-side solution and one translated and executed by \W.

The language-side version of an instrumented \textbf{basicNew} looks as follows: 
\begin{stcode}{}
Class>>basicNew
	| object |
	object := super basicNew.
	FileStream stdout ifNotNil: [ :stream |
		stream << object nbAddress << String lf ].
\end{stcode}
However, printing on the standard output might easily fall in a recursive loop as described before. 
Hence the safe version needs an additional recursion guard:
\begin{stcode}{}
Class>>basicNew
	RecursionGuard
	 	ifStackContains: #basicNew
		do: [ ^ self unmodifiedBasicNew ].
	FileStream stdout ifNotNil: [ :stream |
		stream << self name << String lf ].
	^super basicNew
\end{stcode}
%
\noindent In \W we define a new version of \textbf{basicNew} in a small Slang method which itself is a wrapper around the unmodified primitive: 

\begin{stcode}{} 
WaterFall >> slangBasicNew
	| oop value | 
	oop := self stackAt: 0. 
	self
		callVMFunction: #printOop 
		withArguments: { oop }.
	^self 
		callVMFunction: #primitiveNew 
		withArguments: {}.
\end{stcode}   

Much like the reflective solution, the instrumented Slang version of the primitive delegates the main functionality to the original one.
The reflective version uses a normal message send to call the original primitive. 
In Slang, after calling a low-level function that is responsible for printing the object memory address, it performs a function call to the VM-level \textbf{basicNew} function.




\subsubsection{Results}
\begin{figure}[H]
\begin{center}
	\includegraphics[width=\linewidth]{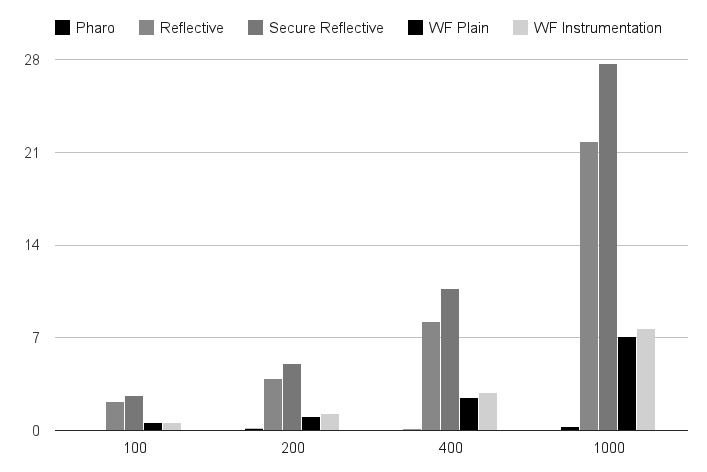} 
	\caption{Instrumenting the creation of objects with reflection and with \W}
	\label{fig:comparisonnew}
\end{center}
\end{figure}

In Figure~\ref{fig:comparisonnew} we compare the run times for different instrumentation approaches of the \textbf{basicNew} primitive. 
We make different comparisons, measuring the creation of 100 and up to 1000 objects. We run the experiment for: the standard primitive, a version with an unsafe reflective instrumentation, a safe version of a reflective instrumentation with a secure guard, a primitive compiled by \W for creating objects and finally an instrumented \W version.
Table~\ref{benchmark-basicnew} illustrates the relative slowdown factors.

Some interesting conclusions can be inferred from the chart.
First of all, it can be seen that in most cases the time for creating objects with the standard \PH primitive is almost negligible. This is because creating objects is an essential task in an object-oriented environment and thus the primitive is highly optimized. Then it is also visible that just the overhead of calling \W is quite high. This is because \W is still a prototype and we have not focused on optimizations yet. Actually, is not our goal to compare \W with a fully optimized C compiler.

It is also worth noting that the cost of instrumentation using \W is negligible compared to the reflective techniques. As the chart exhibits, plain \W versions are similar in time to \W instrumented version while the table shows that reflective instrumentation imposes a slowdown factor of $78\times$ and $99\times$. 
This is encouraging since it shows that if \W is further optimized it would get closer to the pure static times. 

Finally, concerning the analysis only for instrumented versions, the chart and table also exhibit that the reflective solutions are considerably slower than the ones based on \W (a slowdown factor is between $3$ and $4$ depending on the approach).
We compare \W with two reflective approaches for completeness reasons but, since \W avoids the recursive loop, a fair comparison will be with the safer version. 
The final cost of this operation is correlated with the size of the call stack it must traverse. Since this slowdown factor of almost $4\times$ was obtained using the benchmarking framework that generates a pretty shallow call stack, we conjecture that a real application may suffer more overheads, favoring our approach of using  \W. 
Moreover, the implementation in \W of simple optimizations like function inlining will surely enlarge even more the differences.

\subsection{Towards Dynamic Customizable VMs}
In order to provide a stronger validation  we choose an interesting and general enough plug-in, frequently used by most of the users. 
We  basically strip it from the VM. Then, we use \W for dynamically and lazily compiling and executing the functions which are actually used. 

\subsubsection{Experiment}
In \PH all file related operations are delegated to a plug-in named the \textbf{File Plug-ins}.
We choose to experiment with this plug-in since it encompasses several functionalities and is exhaustively used by most standard users. It also is one of the most complex plug-ins, because it has a strong interaction with the OS. 

The experiment consists on evaluating the feasibility of extracting the functionality from VM-side and using \W for generating only the functionality required on demand.  


\subsubsection{Results}
\begin{figure}[t]
\begin{center}
	\includegraphics[width=\linewidth]{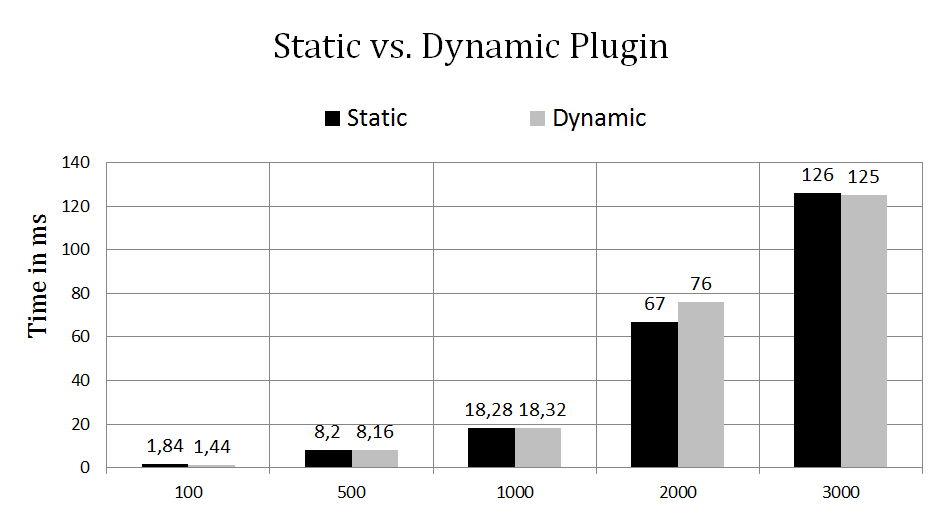} 
	\caption{Static vs. dynamic plug-in comparison for creating files}
	\label{fig:comparisonplugins}
\end{center}
\end{figure}

Following the same methodology already explained for the previous validation, we compare the execution time for creating directories with the standard static plug-in with the time it takes for creating them with the function dynamically generated with \W.
Figure~\ref{fig:comparisonplugins} shows the results.
The X axis represents the number of different directories the test created. Each one is a call to the primitive.
The Y axis exhibits the average time.

It can be clearly observed that concerning the performance is almost the same with the static as with the dynamic approach. Also it is visible, and it seems like a contradiction, that the growth is non-linear. For instance, creating 2000 directories is more than twice than creating 1000. This is because the more directories created, the more the environment works, more objects created and then more garbage collection cycles. 

We noted that we obtained very good timing results. We conjecture that this could be related to the fact that the main cost of file operations is expended by the OS. For other plug-ins, the \W version may expose some performance degradation. However, the main focus of this experiment is to show the feasibility of our approach on removing plug-ins from the VM. \textbf{File Plug-ins} is an excellent candidate for that goal since it is one of the most-used plug-ins. We managed to move it completely to language-side and show that it has no impact on the usability of the environment. The discussion about efficiency was already presented on the previous experiment.



\section{Related Work}
An approach more similar to ours is  QUICKTALK~\cite{Ball86a}. 
It was conceived to compile \ST code directly to binary code. Developers must include type annotations in order to bound all method invocations at compile time. 
This focus on performance and the development of a complex compiler for a new \ST dialect.
In contrast \W creates a bridge between the VM and the language-side. 
Unfortunately, this tool is not available to perform an empirical evaluation.
  

Another noteworthy \ST implementation is Smalltalk/X\footnote{\url{http://www.exept.de/en/products/smalltalkx}} that has excellent C integration built into the language.
Using dedicated syntax, C expressions can be written inline.
Smalltalk/X explicitly uses this approach to define primitives transparently at language-side (in C).
Much like \W primitives can be modified at runtime, however Smalltalk/X does not focus on a one-language approach where the VM and the dynamic primitives share the same language.

High-level low-level programming~\cite{Fram09a} encourages to use high-level languages for system programming.
In this work authors present a low-level framework which is used as system interface for Jikes, an experimental Java VM.  
Methods have to be annotated in order to tell compiler to use the low-level functionality. Although this work is an step forward the use of high-level languages to build system software, the strong separation between low-level code and runtime does not allow for reflective extensions of the runtime.


\PH benefits from Squeak\cite{Inga97a} an early self-hosted VM for \ST.
More recent self-hosted approaches include Klein~\cite{Unga05a} for Self, PyPy \cite{Rigo06a} for Python, Maxine \cite{Wimm13a} for Java or Tachyon~\cite{Chev11a} for JavaScript.
%
The Maxine VM stands out as it truly focuses on productivity and developer interaction.
Maxine uses abstract and high-level representations of VM-level concepts and consistently exposes them throughout the development process.
Inspectors at multiple abstraction levels are readily available while debugging, giving insights to the complete VM state.
Compared to Maxine, \W currently lacks the debugging tools which would enable a truly seamless interaction with the low-level world.
However, Maxine focuses on Java, a language with inferior reflective capabilities compared to \PH.
Hence the live interaction with the VM is only presented in the development phase and not exposed to the language-side.
Maxine would be an excellent candidate to implement our approach for Java.

Apart from the VM community, \W shares many ideas with research conducted in the high-level reflection domain.
For instance, Röthlisberger et al.~\cite{Roet07b} present a tool to mitigate runtime adaptions with unanticipated partial behavioral reflection .
Built on top of \ST they present a tool to limit the computational overhead that dynamic runtime reification introduces.
This contrast with Aspect Oriented Programming solutions where the runtime has to be modified upfront to allow for partial behavior reflection.
In the scope of our work, the latter approach corresponds to static changes at VM-level.
\W on the other hand allows for unanticipated changes.

\section{Conclusions}
In this work we present a toolchain that allows for altering and configuring components such as primitives and plug-ins at runtime. 
The main component is \W, a dynamic and reflective translator for Slang, a restricted subset of \ST.
\W is implemented completely at language-side and it is integrated in \PH. 
\W generates primitives and plug-ins on demand and executes them on the fly.
 
Even though \W still lacks substantial optimizations it outperforms dynamic instrumentation of primitives based on purely reflective approaches. It also
enables to have dynamically generated plug-ins which perform reasonably well with respect to their statically compiled counterparts.  

We believe this approach provides a flexible mechanism for adapting and evolving VMs and enables developers to deploy them with a smaller footprints. VMs can be later customized by users according to their needs, at runtime and without resorting to external tools.






Concerning conceptual improvements, with \W we encourage the use of high-level low-level programming at runtime.
Furthermore we advocate the importance of self-hosted VMs to control and modify essential parts of them from language-side.

\cb{I do not parse...}
We also managed to enhance quality properties of the environment by decoupling it from the low-level building infrastructure while modifying low-level behavior. 
Operating systems, compilers and linkers tends to impose important constraints.
For instance, with \W it can be completely avoided for altering some components at runtime the common low-level cycle: compile $\rightarrow$ link $\rightarrow$ save to file $\rightarrow$ load to memory.
 As a consequence, our approach also improves portability since the compilation infrastructure is in general very dependent on the platform.

 \subsection{Perspectives}
 
Even current self-hosted or metacircular VMs freeze many aspects at compile-time.
We think VMs should be easier to evolve and adapt dynamically, using reflective runtime capabilities.  
At the same time, the VM performance should be comparable with the solutions written in low-level languages. 
We believe the work presented in this paper together with other techniques such as gradual typing and type inference, opens up new perspectives about the possibility of approaching our vision.

In this setting we plan to work on improving the \W back-end in order to produce more efficient code. This includes powerful static and dynamic analysis techniques targeted specially for highly dynamic environments. 
 






\bibliographystyle{abbrv}

\bibliography{local}



\end{document}

%% file: header.tex
\usepackage[utf8]{inputenc}
\usepackage[T1]{fontenc}
\usepackage{graphicx}
\graphicspath{{figures/}}

\usepackage{xspace}
\usepackage[htt]{hyphenat} 
\usepackage{alltt} 
\usepackage{amssymb}
\usepackage{amsmath}
\usepackage{textcomp}
\usepackage{booktabs} 
\usepackage{subfigure}
\usepackage{pifont}
\usepackage{xspace}
\usepackage{lmodern}
\usepackage{natbib}
\usepackage{url}
\usepackage{enumitem} 	
\usepackage{stmaryrd}   
\usepackage{float}      
\usepackage{flushend}   
\usepackage[hypcap]{caption} 
\DeclareCaptionType{copyrightbox}
\usepackage[normalem]{ulem} 
\usepackage{xcolor}

\ifthenelse{\boolean{showcomments}}
{
	\newcommand{\ugh}[1]{\textcolor{red}{\uwave{#1}}}   
	\newcommand{\del}[1]{\textcolor{red}{\sout{#1}}}    
}{
	\newcommand{\ugh}[1]{#1}                            
	\newcommand{\del}[1]{}                              
	
}
\usepackage{amssymb}

\ifthenelse{\boolean{showcomments}}{
	\newcommand{\nbc}[3]{
		{\colorbox{#3}{\bfseries\sffamily\scriptsize\textcolor{white}{#1}}}
 		{\textcolor{#3}{\sf\small$\blacktriangleright$\textit{#2}$\blacktriangleleft$}}}
}{
	\newcommand{\nbc}[3]{}
	
}

\usepackage{needspace}
\newcommand{\needlines}[1]{\Needspace{#1\baselineskip}}

\usepackage{xcolor}
\definecolor{source}{gray}{0.95}
\usepackage{listings}
\lstdefinelanguage{ST}{
    keywordsprefix=\#,
    morekeywords=[0]{true,false,nil},
    morekeywords=[1]{self,super,thisContext},
    morekeywords=[2]{ifTrue:,ifFalse:,whileTrue:,whileFalse:,and:,or:,xor:,not:,by:,timesRepeat:},
    sensitive=true,
    morecomment=[s]{"}{"},
    morestring=[d]',
    escapechar={!},
    alsoletter={., :, -, =, +, <},
    moredelim=**[is][\itshape]{/+}{+/},
    literate=
        {^}{{$\uparrow$}}1
        {:=}{{$\leftarrow$}}1
        {~}{{$\sim$}}1
        {-}{{\sf -\hspace{-0.13em}-}}1  
        {+}{\raisebox{0.08ex}{+}}1		
        , 
    style=STStyle
}
\lstdefinestyle{STStyle}{
    tabsize=4,
    basicstyle=\ttfamily\small,
    keywordstyle=\bf\ttfamily,
    stringstyle=\mdseries\slshape,
    commentstyle=\it\rmfamily\color{darkgray}, 
    commentstyle=\mdseries\slshape\color{gray},
    emphstyle=\bf\ttfamily,
    escapeinside={!}{!},
    keepspaces=true
} 

\lstnewenvironment{stcode}    [2][]{\lstset{language=ST,#1}\needlines{#2}}{}
\lstnewenvironment{ccode}     [2][]
    {\lstset{language=C,numbers=left,escapechar=\$,numberstyle=\tiny,#1}\needlines{#2}}{}

\lstnewenvironment{numstcode} [2][]
    {\lstset{language=ST,numbers=left,numberstyle=\tiny,numbersep=2pt,#1}\needlines{#2}}{}
\lstnewenvironment{numstcodecont} [2][]
    {\lstset{language=ST,numbers=left,numberstyle=\tiny,numbersep=2pt,firstnumber=last#1}\needlines{#2}}{}

\lstnewenvironment{code}{%
	\lstset{%
		frame=single,
		framerule=0pt,
		mathescape=false
	}
}{}

\newcommand{\ttt}[1]{\texttt{#1}}

\newcommand{\seclabel}[1] {\label{sec:#1}}

\newcommand{\secref}  [1] {\ref{sec:#1}}

\renewcommand{\eqref} [1] {\ref{eq:#1}}
\newcommand{\ST}  {Small\-talk\xspace}

\newcommand{\PH}  {Pharo\xspace}

\usepackage[
    colorlinks=true, 
    urlcolor=black, 
    linkcolor=black, 
    citecolor=black, 
    bookmarksnumbered=true, 
    bookmarks=true]{hyperref}